\documentclass[%
reprint,
superscriptaddress,
amsmath,amssymb,
aps,
prl,
]{revtex4-1}
\usepackage{graphicx}
\usepackage{dcolumn}
\usepackage{bm}
\usepackage{amsmath}
\usepackage{hyperref}
\hypersetup{colorlinks=true,linkcolor=blue,anchorcolor=blue,citecolor=blue,urlcolor=blue}
\usepackage{lineno}
\usepackage{enumerate}
\usepackage{color}

\begin{document}
\preprint{APS/123-QED}

\title{Ion Kinetics and Neutron Generation Associated with Electromagnetic Turbulence in Laboratory-scale Counter-streaming Plasmas}

\author{P. Liu}
\affiliation{Institute for Fusion Theory and Simulation, School of Physics, Zhejiang University, Hangzhou 310058, China}%

\author{D. Wu}
\email{dwu.phys@sjtu.edu.cn}
\affiliation{Key Laboratory for Laser Plasmas and School of Physics and Astronomy, Collaborative Innovation Center of IFSA (CICIFSA), Shanghai Jiao Tong University, Shanghai 200240, China}%

\author{T. X. Hu}
\affiliation{Institute for Fusion Theory and Simulation, School of Physics, Zhejiang University, Hangzhou 310058, China}%
\affiliation{Key Laboratory for Laser Plasmas and School of Physics and Astronomy, Collaborative Innovation Center of IFSA (CICIFSA), Shanghai Jiao Tong University, Shanghai 200240, China}%

\author{D. W. Yuan}
\affiliation{Key Laboratory of Optical Astronomy, National Astronomical Observatories, Chinese Academy of Sciences, Beijing 100012, China}%

\author{G. Zhao}
\affiliation{Key Laboratory of Optical Astronomy, National Astronomical Observatories, Chinese Academy of Sciences, Beijing 100012, China}%

\author{Z. M. Sheng}
\email{zmsheng@zju.edu.cn}
\affiliation{Institute for Fusion Theory and Simulation, School of Physics, Zhejiang University, Hangzhou 310058, China}%

\author{X. T. He}
\affiliation{Institute for Fusion Theory and Simulation, School of Physics, Zhejiang University, Hangzhou 310058, China}%

\author{J. Zhang}
\affiliation{Key Laboratory for Laser Plasmas and School of Physics and Astronomy, Collaborative Innovation Center of IFSA (CICIFSA), Shanghai Jiao Tong University, Shanghai 200240, China}%

\date{\today}

\begin{abstract}
Electromagnetic turbulence and ion kinetics in counter-streaming plasmas hold great significance in laboratory astrophysics, such as turbulence field amplification and particle energization.
Here, we quantitatively demonstrate for the first time how electromagnetic turbulence affects ion kinetics under achievable laboratory conditions (millimeter-scale interpenetrating plasmas with initial velocity of $2000\ \mathrm{km/s}$, density of $4 \times 10^{19}\ \mathrm{cm}^{-3}$, and temperature of $100\ \mathrm{eV}$) utilizing a recently developed high-order implicit particle-in-cell code without scaling transformation.
It is found that the electromagnetic turbulence is driven by ion two-stream and filamentation instabilities. For the magnetized scenarios where an applied magnetic field of tens of Tesla is perpendicular to plasma flows, the growth rates of instabilities increase with the strengthening of applied magnetic field, which therefore leads to a significant enhancement of turbulence fields. Under the competition between the stochastic acceleration due to electromagnetic turbulence and collisional thermalization, ion distribution function shows a distinct super-Gaussian shape, and the ion kinetics are manifested in neutron yields and spectra. Our results have well explained the recent unmagnetized experimental observations, and the findings of magnetized scenario can be verified by current astrophysical experiments.
\end{abstract}

\maketitle
The electromagnetic instabilities, which usually arise from the interpenetrating plasma systems of the ejections from supernova explosions and the surrounding interstellar medium, have become one of long-standing research focuses in laboratory astrophysics communities.
It is believed to be the mechanism driving collisionless shocks and accelerating particles in several astrophysical events, such as supernova remnants \cite{MiceliP2019} and gamma-ray bursts \cite{PruetPhysRevD2004}. In such so-called collisionless systems, the mean-free-path of Coulomb collision $\lambda_{mfp}$ is larger than the scale of interest $ L $, allowing the plasma flows to interpenetrate each other. It is noteworthy that particle motions can be modified by self-generated electromagnetic fields, originating from the Biermann battery effect \cite{WidrowRevModPhys2002} or Weibel-type instabilities \cite{WeibelPhysRevLett1959,Fried1959}, which can also affect the nuclear reaction process of charged particles.

\begin{figure}[t]
  \includegraphics[scale=0.56]{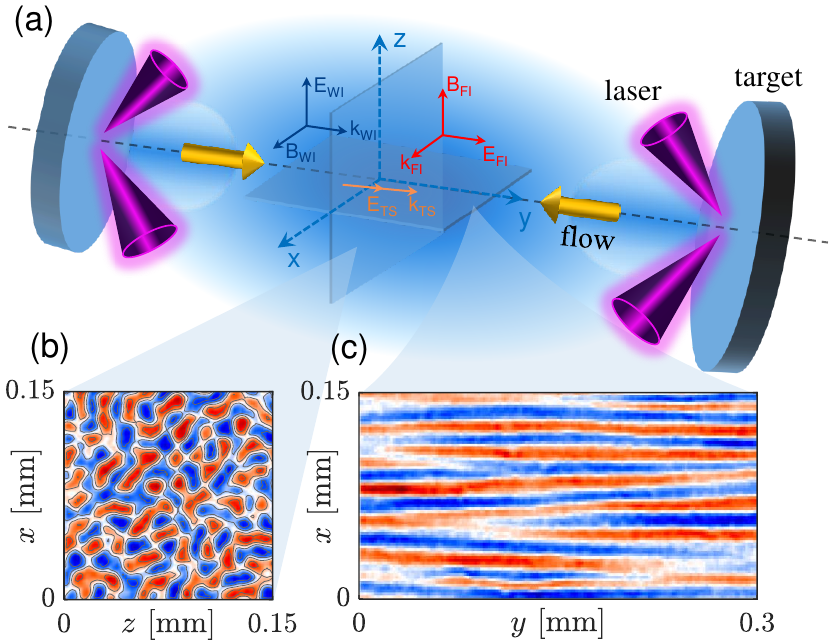}
  \caption{\label{fig:figure1} (a) Schematic diagram of two counter-streaming plasmas generated by laser-ablation targets, there are three possible unstable modes: the filamentation mode (labeled with ``FI'') with $ \mathbf{k}\perp \mathbf{v}_0 $, Weibel mode (labeled with ``WI'') with $ \mathbf{k} \parallel \mathbf{v}_0 $, and two-stream mode (labeled with ``TS'') with $ \mathbf{k} \parallel \mathbf{v}_0 $, where $ \mathbf{v}_0 $ is the flow velocity. Maps of filamentous currents from PIC simulations in $z$-$x$ (b) and $y$-$x$ (c) planes.}
\end{figure}

In fact, although the growth of plasma instabilities is limited by collisional dissipations in experiments \cite{RyutovPoP2014}, the electromagnetic field structures are successfully observed in laser-driven counter-streaming plasmas \cite{KuglandNP2012,RossPhysRevLett2013,FoxPhysRevLett2013,HuntingtonNP2015,ParkPoP2015,HuntingtonPoP2017,RossPhysRevLett2017,FiuzaNP2020,HigginsonPOP2019}.
Figure \ref{fig:figure1}(a) shows the schematic diagram, in which counter-streaming plasmas are produced with lasers irradiating two separated targets, and nuclear reactions occur when two plasma jets collide. Such a counter-streaming plasma system provides an important platform to investigate the particle temperature coupling \cite{RossPhysRevLett2013}, collisionless shock \cite{ParkPoP2015,  FiuzaNP2020}, electron acceleration \cite{FiuzaNP2020}, and neutron generation \cite{RossPhysRevLett2017,HigginsonPOP2019}. For the plasma flows with relatively high density and velocity, both thermonuclear ($ \lambda_{mfp} \leq L $) and beam-beam ($ \lambda_{mfp} \geq L $) reactions are usually included.
The thermonuclear neutrons are usually produced by thermal collisions when the plasmas are in equilibrium, i.e., $ \lambda_{mfp} \leq L $, such as DD fusion reactions in interpenetrating plasmas of CD/CH case.
While for CD/CD scenario, neutron generations are dominated by the counter-streaming CD/CD flows, namely beam-beam reaction, which is a non-equilibrium process and can occur when $ \lambda_{mfp} \geq L $ \cite{HigginsonPOP2019}.
Recent studies showed that the discrepancies in experimental observations and simulation results for neutron yields \cite{RossPhysRevLett2017} and velocities \cite{HigginsonPOP2019} are considered to be likely attributed to electromagnetic effects, which are produced by plasma instabilities \cite{KuglandNP2012,RossPhysRevLett2013,FoxPhysRevLett2013,HuntingtonNP2015,HuntingtonPoP2017,ParkPoP2015,FiuzaNP2020}. To the best of our knowledge, distinguishing the effects of electromagnetic instabilities on ion kinetics in counter-streaming plasmas is challenging for experiments. Besides, laboratory-scale particle-in-cell (PIC) simulations including electromagnetic turbulence, collisional thermalization, and neutron diagnosis are not yet available in the publications due to the huge computational burden. Therefore, the influences of electromagnetic instability and turbulence on ion kinetics and neutron generation remain largely ambiguous.

In this Letter, we utilize a recently developed high-order implicit PIC code to quantitatively investigate how electromagnetic turbulence affects ion kinetics under realistic laboratory conditions. Supported by neutron diagnostics, including neutron yields and spectra, we demonstrate that the electromagnetic turbulence is driven by both ion two-stream instability (ITSI) and ion filamentation instability (IFI). Furthermore, in magnetized scenarios, the growth rates of these instabilities increase with the strengthening of the applied magnetic field, leading to a significant enhancement of turbulence fields and the emergence of super-Gaussian shapes in the ion distribution functions. This strongly be associated with turbulence field amplification \cite{BohdanPRL_TurbAmp,Bell_TrubAmp,PetersonPRL2021,Peterson_2022} and particle acceleration \cite{Comisso_PRL2018,ZhdankinPRL2017,Petrosian_2012,ZhdankinPRL2019}. Additionally, following the definitions of unstable modes provided by Bret and Lazar $et\ al$. \cite{BretPhysRevLett2004,Lazar_2009}, we systematically derive and discuss various plasma instabilities, including ITSI, IFI, ion Weibel instability (IWI), and drift-kink instability (DKI).

A full linear kinetic model allows us to self-consistently account for ITSI, IFI, IWI, and DKI in the unmagnetized counter-streaming systems, and the detailed derivations of dispersion functions are presented in Sec. I of the Supplemental Material \cite{Supplement}. It is found that the ion Weibel mode is stable, and when considering the timescale of full development of instabilities (i.e., several hundred picoseconds), the DKI can be negligible due to its characteristic growing-up time of $\sim$7 ns.
\textcolor{blue}{We find that the electrostatic ITSI is an oblique instability mode \cite{BretPoP2010,kato_2010,Skoutnev_2019} with growth rate of $ \Gamma _{\mathrm{ITSI}}\simeq 2.2 \times 10^{-3}\omega _{pe} $}, and its characteristic growing-up time is $\sim$1 ps, which can provide a possible seed for the development of turbulence fields; the IFI with growth rate of $ \Gamma _{\mathrm{IFI}}\simeq 4.3 \times 10^{-3}\omega _{pi} $ and growing-up timescale of $\sim$100 ps is another factor to drive electromagnetic turbulence, and its growth rate depends on the anisotropy of ion beams, given by \cite{Supplement}
\begin{equation}\label{Eq1}
  A_{i}=\frac{2v_{0}^{2}+v_{ti ,\parallel}^{2}}{v_{ti ,\perp}^{2}}-1,
\end{equation}
indicating that $A_i$ not only depends on the thermal anisotropy but also on ratio of flow velocity to thermal velocity, where $v_{ti ,\parallel}$ and $v_{ti ,\perp}$ are respectively the parallel and perpendicular thermal velocities of ions with respect to flow velocity $v_{0}$, $\omega_{pe}$ and $\omega_{pi}$ are the plasma frequencies of electron and ion, respectively.
In such counter-streaming systems, $ A_i $ mainly depends on $2v_0^2/v_{ti,\perp}^2$ for most recent experiments  \cite{KuglandNP2012,RossPhysRevLett2013,FoxPhysRevLett2013,HuntingtonNP2015,ParkPoP2015,HuntingtonPoP2017,RossPhysRevLett2017,FiuzaNP2020,HigginsonPOP2019} due to $ v_0\gg v_{ti} $ at the linear stage, which have been validated by our numerical solutions \cite{Supplement}.
Note that $ A_i $ is only determined by thermal anisotropy when $v_0 \rightarrow 0$, reducing to classical Weibel mode \cite{WeibelPhysRevLett1959}.

Compared to the linear regime, the nonlinear processes usually play a more important role, and in fact, the nonlinear electromagnetic turbulence and its influence on ion kinetics, which depend on both the plasma flow and ambient medium conditions, are far from completely being understood.
Fortunately, $ab\ initio$ PIC simulations have significantly improved our ability to study the nonlinear processes associated with turbulence fields and ion kinetics.
Here, a series of large-scale two-dimensional (2D) PIC kinetic simulations are performed by employing the LAPINS code \cite{Wu2019,Wu2021}, which is based on a high-order implicit algorithm, eliminating the numerical cooling found in the standard implicit PIC methods by the use of pseudo-electric-field \cite{Wu2019}. Recently, a pairwise nuclear reaction module was developed in the LAPINS code along with the binary collision algorithm  \cite{Wu2021},
it thus possesses the capability to self-consistently study kinetic instabilities, collisional thermalizations, and nuclear reactions \cite{Wu2019,Liu_2022}. It should be remarked that the scaling transformation PIC method \cite{HuntingtonNP2015,HuntingtonPoP2017,FiuzaNP2020}, in which artificially scaling up beam velocity and reducing ion-to-electron mass ratio, is no longer applicable because the collisional thermalization and nuclear reaction cannot be achieved. Moreover, the effect of nonphysical mass ratio on ion filamentation dynamics should not be negligible at the nonlinear stage \cite{RuyerPhysRevLett2016}.

\begin{figure*}[t]
  \includegraphics[scale=0.4]{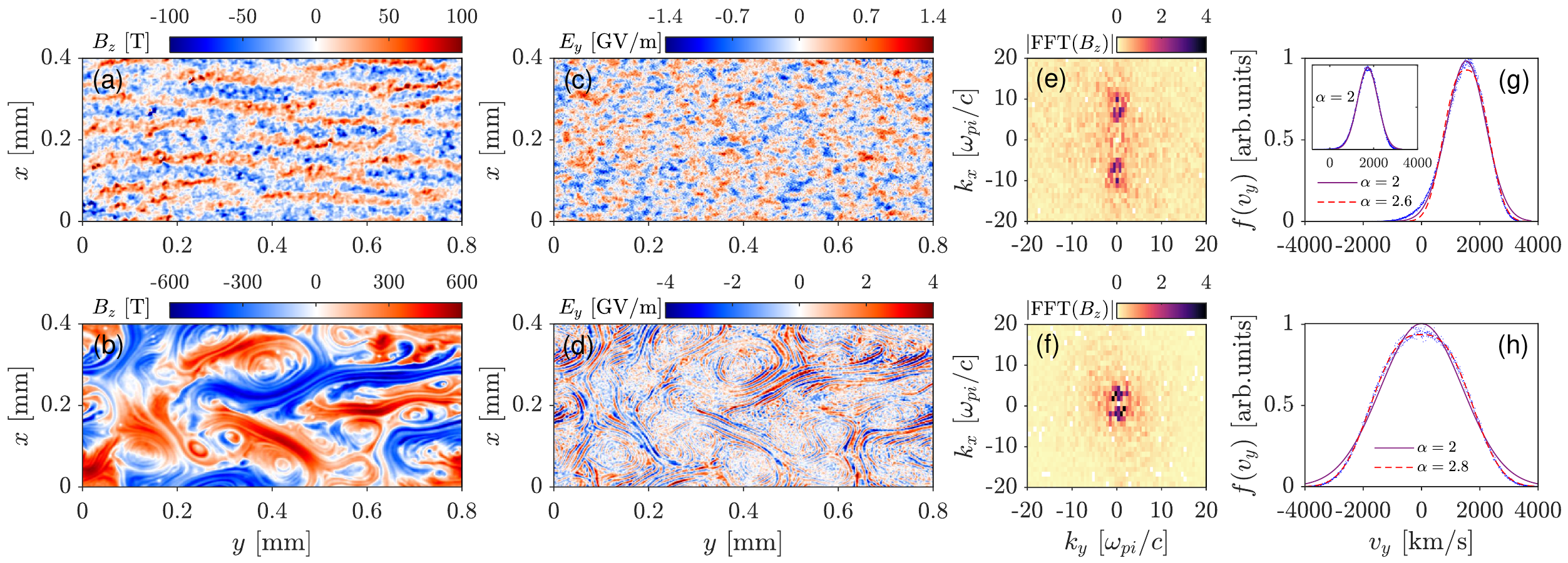}
  \caption{\label{fig:figure2} Top: unmagnetized case with $B_0 = 0$ T. Bottom: magnetized case with $B_0 = 50$ T. Snapshots of magnetic field (a),(b) and electric field (c),(d) at $t = 800$ ps. (e) and (f) are 2D Fourier transforms of magnetic field shown in (a) and (b), respectively. (g),(h) Distribution functions of the deuteron beams at  $t = 6$ ns, where the blue dots and solid (dashed) lines respectively represent the simulation data and Gaussian (super-Gaussian) fitting curves with the form of $\exp \left[-(v_y/v_{th})^\alpha \right]$ for high-energy ions. \textcolor{blue}{The inset of (g) for the case without considering electromagnetic effects, i.e., the electromagnetic fields are not applied to the particles.} We also perform another magnetized case where $B_0 = 10$ T, and the simulation results about the distributions of electromagnetic fields, Fourier transform, and velocity distribution function can be seen in the Supplemental Material \cite{Supplement}}
\end{figure*}

In our simulations, the uniform $ \mathrm{CD}/ \mathrm{CH} $ and $ \mathrm{CD} / \mathrm{CD} $ plasma flows ($ n_\mathrm{C}:n_\mathrm{D} = n_\mathrm{C}:n_\mathrm{H} = 1:2 $), with a realistic ion mass ratio (e.g., $m_\mathrm{D}/m_e=3672$ for $\mathrm{D}$ and $m_\mathrm{H}/m_e=1836 $ for $\mathrm{H}$), propagate along the $\pm y$ directions. The beams follow the typical experimental parameters of NIF and OMEGA with initial velocity of $v_0 = 2000\ \text{km/s}$, density of $n_e = 4 \times 10^{19}\ \text{cm}^{-3}$, and temperature of $T = 100 \ \mathrm{eV}$ \citep{KuglandNP2012,RossPhysRevLett2013,FoxPhysRevLett2013,HuntingtonNP2015,ParkPoP2015,HuntingtonPoP2017,RossPhysRevLett2017,FiuzaNP2020,HigginsonPOP2019}.
The plasma flows are magnetized with a uniform magnetic field $\mathbf{B}_{\mathrm{ext}}=B_0\mathbf{\hat{e}}_x$, and the corresponding initial magnetization is $\sigma _0=B_{0}^{2}/ \left( \mu _0 \sum_i{n_i}m_iv_{i}^{2} \right) \simeq 10^{-3}$ \cite{Demidem_2023}, which is available for current laboratories \cite{Fujioka2013,Law2016}, where $\mu_0$ is vacuum permeability, $n_i$ and $v_i$ are number density and velocity of ion beams, respectively.
The simulation window of sizes $ 0.8 \ \mathrm{mm} \ (y) \times 0.4 \ \mathrm{mm} \ (x) $ and $ 0.4 \ \mathrm{mm} \ (z) \times 0.4 \ \mathrm{mm} \ (x) $ for different configurations with periodic boundary conditions for both particles and fields are adopted. The timestep of 3.3 fs and grid size of $2.0\ \mu$m ($\simeq 0.02 c/\omega_{pi} \simeq 2 c/\omega_{pe}$) with 528 particles per cell are used, where $c$ is light speed in vacuum.

To get a deeper insight of electromagnetic turbulence at the nonlinear stage, we plot the magnetic field $B_z$ and electric field $E_y$ from the unmagnetized case in Figs. \ref{fig:figure2}(a) and \ref{fig:figure2}(c), which show typical turbulence distributions at a nonlinear moment $t = 800$ ps. The average turbulent magnetic field induced by the IFI is on the same order of magnitude as that predicted by magnetic trapping theory \cite{Supplement,Davidson1972}
\begin{equation}\label{Eq2}
   B_{\mathrm{sat}} \simeq \frac{\Gamma_{\mathrm{IFI}}^2m_ic}{q_iv_0k_{\mathrm{sat}}} \simeq 40 \ \mathrm{T},
\end{equation}
where $k_{\mathrm{sat}}$ is the fastest-growing wave number. This nonlinear electromagnetic turbulence is closely related to the plasma instabilities in the counter-streaming system, including both the ITSI and IFI.

It should be noted that for the simulations of the $z$-$x$ plane, the longitudinal ITSI can be excluded, only the transverse IFI develops, and the saturated magnetic field can be well predicted by Eq. (\ref{Eq2}). The simulation results obtained from this configuration are also presented in Supplemental Material \cite{Supplement}.

PIC Simulation results indicate that the turbulence emerges spontaneously from the nonlinear evolution of instabilities. In the magnetized scenario, the electromagnetic turbulence with vortex structures is observed in Figs. \ref{fig:figure2}(b) and \ref{fig:figure2}(d), and turbulence field amplification shows a significant transformation of kinetic energy into waves. 
The average turbulent magnetic fields can reach $\sim$90 T for $B_0 = 10$ T and $\sim$200 T for $B_0 = 50$ T, and corresponding magnetization $ \sigma_1 =B_{1}^{2}/ \left( \mu _0 \sum_i {n_im_iv_{i}^{2}} \right) $ for self-generated field $B_1$ are $\sim$0.0401 and $\sim$0.113, respectively, which are more than 1 order of magnitude higher than that in the unmagnetized case ($\sigma_1 \simeq 0.0028$).
Moreover, we also derive the dispersion relations of the IFI and ITSI for the magnetized counter-streaming plasmas in Sec. III of Supplemental Material \cite{Supplement}, which indicates that the linear growth rates of both are enhanced with the increase of applied magnetic field $ \mathbf{B_{\mathrm{ext}}} $. Consequently, the saturated turbulent magnetic fields are amplified due to $ B_{\mathrm{sat}} \propto \Gamma^2 $.
This electromagnetic turbulence is also analyzed in the spectral space, and in Figs. \ref{fig:figure2}(e) and \ref{fig:figure2}(f), we show the amplitude of Fourier transforms of Figs. \ref{fig:figure2}(a) and \ref{fig:figure2}(b). The turbulent magnetic field has a significant anisotropy distribution with a wave number peaked at $k_m = \left( k_x^2+k_y^2 \right)^{1/2} \simeq 7\omega_{pi}/c$ in the unmagnetized case. While for the magnetized scenario, the turbulence field oscillations with a large wavelength ($ \propto 1/k_m$) are observed, and corresponding wave number drops to $k_m\simeq 3 \omega_{pi}/c$ [see Fig. \ref{fig:figure2}(f)], and it tend to exhibit isotropic turbulence \cite{Kolmogorov1941}.

\begin{figure}[t]
  \includegraphics[scale=0.6]{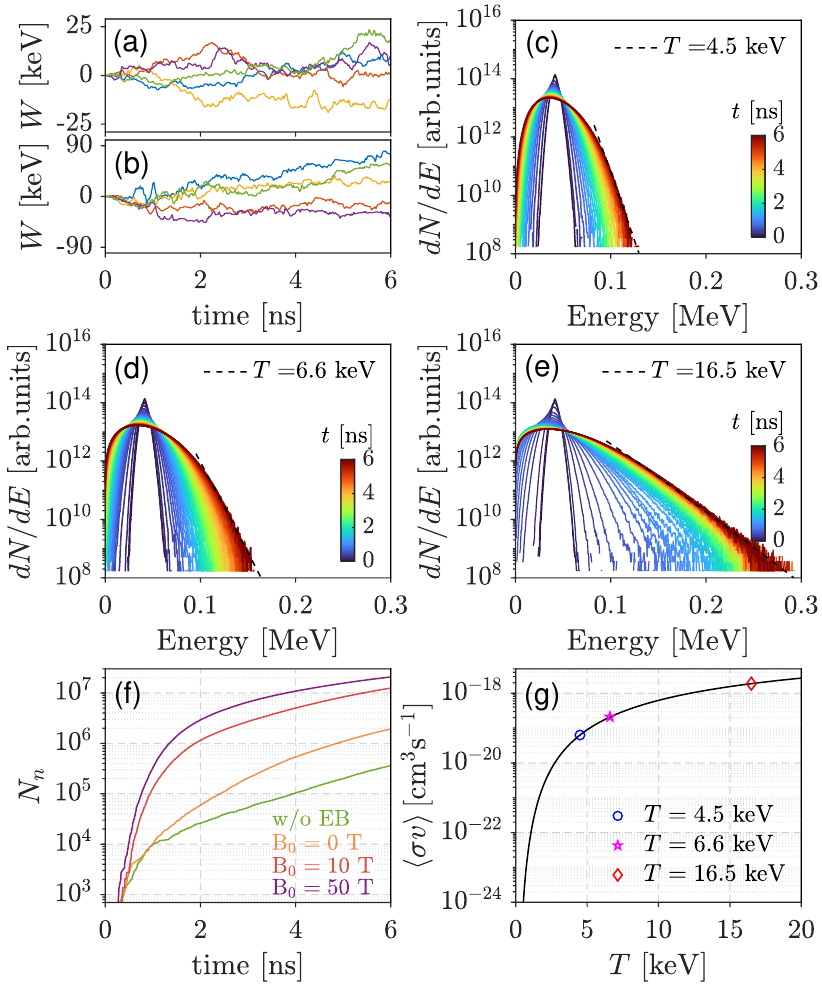}
  \caption{\label{fig:figure3} Temporal evolution of work done by turbulent electric field for sampled deuterons in the unmagnetized case with $B_0 = 0$ T (a) and magnetized case with $B_0 = 50$ T (b), where the colored curves represent different tracked deuterons.
  Temporal evolution of the deuteron energy spectra (colored by time) for three scenarios of excluding electromagnetic field effects (w/o EB) (c), $B_0 = 0$ T (d), and $B_0 = 50$ T (e). (f) Neutron yields as the functions of time in the $ \mathrm{CD} / \mathrm{CH} $ case.
  (g) Average reactivity of DD thermonuclear fusion as the function of temperature \cite{Hively_1977,Atzeni_2004}, where the marks of blue circle, magenta pentagram, and red diamond, with $6.31 \times 10^{-20}$, $2.20 \times 10^{-19}$, and $1.88 \times 10^{-18} \ \mathrm{cm^3s^{-1}}$, correspond to three temperatures in (c)-(e), respectively.}
\end{figure}

To illustrate the influences of such electromagnetic turbulence on ion kinetics, we analyze the velocity distributions and energy spectra of deuterons. As shown in Fig. \ref{fig:figure2}(g), the ion distribution function features a non-Maxwellian profile, and for the high-velocity tail, the falloff is steeper than a usual Maxwellian shape, which is fitted by a super-Gaussian function of $\exp \left[-(v_y/v_{th})^\alpha \right]$ with $\alpha \simeq 2.6$, where $v_{th}$ is a constant. Moreover, driven by strong turbulence fields in the magnetized case, the distribution function is severely deformed with a exponential factor of $\alpha \simeq 2.8$, as illustrated in Fig. \ref{fig:figure2}(h). 
This distinct super-Gaussian shape is a result of the competition between stochastic acceleration due to electromagnetic turbulence and collisional thermalization. Note that similar super-Gaussian distributions of electrons are observed due to inverse bremsstrahlung absorption in laser-plasma interactions \cite{Matte_1988,Non-MaxwellianPRL}. As expected, when the electromagnetic effects are not taken into account, a standard Maxwellian distribution with $\alpha \simeq 2$ appears in the inset of Fig. \ref{fig:figure2}(g).
Fig. \ref{fig:figure3}(a) plots the work done by turbulent electric field on deuterons, which shows that the deuterons experience stochastic acceleration or deceleration in turbulence fields, and some of them obtain energies of about 25 keV, leading to a broadened energy spectrum in the unmagnetized case, compared with the case where the electromagnetic effects are ignored, as shown in Figs. \ref{fig:figure3}(c) and \ref{fig:figure3}(d).
By fitting the high-energy tail of ion spectra, the ion temperatures in these two cases are $ T \simeq 4.5 $ keV and $6.6$ keV, respectively.
Furthermore, in the magnetized case with $B_0=50 \ \mathrm{T}$, the deuteron beams can be further heated with a temperature of $ T \simeq 16.5 $ keV and cutoff energy of  $ E_c \simeq 0.29 $ MeV due to the work done by a stronger turbulent electric field [see Figs. \ref{fig:figure3}(b) and \ref{fig:figure3}(e)].

\begin{figure}[t] 
  \includegraphics[scale=0.59]{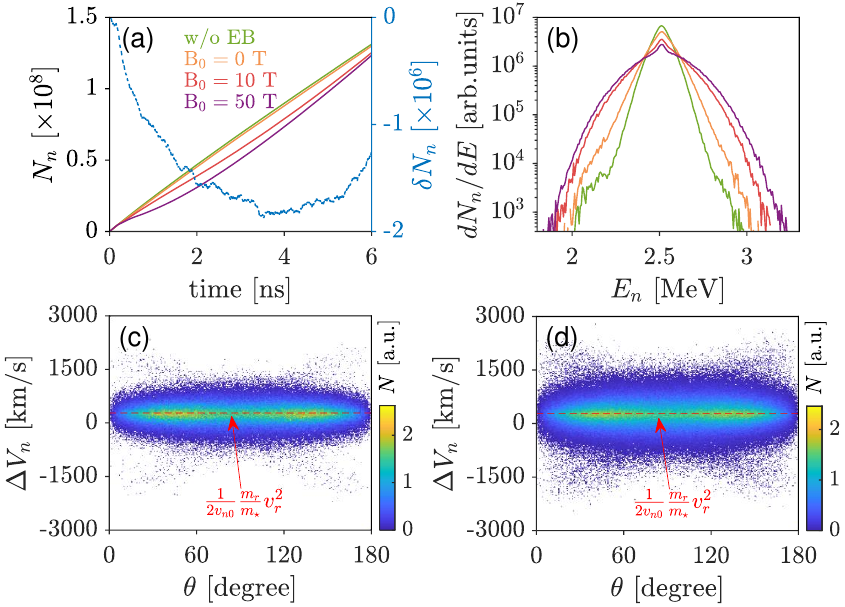}
  \caption{\label{fig:figure4} Temporal evolution of neutron yields (a) and energy spectra at $t=6$ ns (b) in different $ \mathrm{CD}/ \mathrm{CD} $ cases of without considering electromagnetic effects (w/o EB), $B_0=0$ T, $B_0=10$ T, and $B_0=50$ T. The blue line in (a) represents the yield difference between the cases of w/o EB and $B_0 = 0\ \mathrm{T}$ as the function of time. Neutron velocity shifts as the function of emitted angle $\theta$ in the cases of w/o EB (c) and $B_0 = 0\ \mathrm{T}$ (d), where the red dashed lines represent $\Delta V_n = m_r v_r^2/(2v_{n0} m_\star)$, $v_{n0} $ is the velocity of a neutron created from a cold and stationary deuterium plasma, $m_r/m_\star \simeq 0.748$ and $v_r = 2v_0$.}
\end{figure}

Compared with the low-energy deuterons, the high-energy ones corresponding to larger nuclear reaction cross sections have a greater contribution to DD reactions $ \mathrm{D}(\mathrm{d},\mathrm{n}) ^3\mathrm{He}\ \left( Q=3.269\ \mathrm{MeV} \right)$ \cite{HigginsonPOP2019,Atzeni_2004}.
In the unmagnetized $ \mathrm{CD} / \mathrm{CH} $ case, the turbulence fields result in an about fivefold enhancement of neutron yield [see Fig. \ref{fig:figure3}(f)], which provides an explanation for a confusing point that was not fully clarified in the work of Ross $et \ al.$ \cite{RossPhysRevLett2017}, where the experimentally observed neutron yield is larger than that obtained from the simulation without considering electromagnetic effects by a factor of $\sim$6.
When considering pure thermonuclear reactivities \cite{Hively_1977,Atzeni_2004}, we note that the neutron yield is expected to increase by approximately 3.5 times due to the electromagnetic effects [see Fig. \ref{fig:figure3}(g)], which indicates that the neutron yield originates not only from thermonuclear reactions but also from the beam-beam reactions triggered by $ \sim $2.4\% reversed deuterons [see Fig. \ref{fig:figure2}(g)] due to the electromagnetic turbulence.
For the magnetized case where $ B_0 = 50 \ \mathrm{T} $, $N_n \simeq 2.1 \times 10^7$, which is about 1 order of magnitude higher than that in the unmagnetized case [see purple line in Fig. \ref{fig:figure3}(f)].

For $ \mathrm{CD} / \mathrm{CD} $ case, the thermonuclear reaction process is very similar to that in $ \mathrm{CD} / \mathrm{CH} $ case. However, the difference of neutron yield caused by electromagnetic turbulence fields becomes more complex because the beam-beam reactions from different flows are dominant.
\textcolor{blue}{When $ t \gtrsim 0.15 $ ns, the yield difference $\delta N_n$ between the cases of w/o EB and $B_0 = 0\ \mathrm{T}$ is rapidly enlarged with $ \left| \delta N_n \right| \sim 10^6 $}, as shown in Fig. \ref{fig:figure4}(a), which corresponds to the linear period of the IFI growth. This is explained as follows, as illustrated in Figs. \ref{fig:figure1}(b) and \ref{fig:figure1}(c), we witness the well-defined ion filamentous currents at the linear stage, showing a self-organized misalignment distribution between two ion beams, which can diminish the beam-beam reactivity \cite{Supplement}, and a significant drop can be observed in the magnetized cases due to the enhanced instabilities [see red and purple lines in Fig. \ref{fig:figure4}(a)].
When $t \gtrsim 4$ ns, $\delta N_n$ starts to increase, indicating that the electromagnetic turbulence plays an increasingly important role in ion energization at the nonlinear stage. 

Additionally, a striking distinction in the neutron spectra is clearly visible in Fig. \ref{fig:figure4}(b), in which the full width at half maximum (FWHM) of energy spectrum from the case of $B_0=0 \ \mathrm{T}$ is $\Delta E_n \simeq 0.142 \ \mathrm{MeV} $, which is $\sim$34\% larger than that in the case excluding electromagnetic effects, where $\Delta E_n \simeq 0.106 \ \mathrm{MeV}$. 
This is because the turbulence fields create greater center-of-mass velocity $v_{cm}$ and relative velocity $ v_r $, which broadens the neutron velocity shifts  $v_n - v_{n0} \equiv \Delta V_n$, formulated by \cite{HigginsonPOP2019}
\begin{equation}\label{Eq3}
 \Delta V_n   \simeq  v_{cm}\cos \theta - \frac{v_{cm}^2}{2v_{n0}} \sin^2\theta + \frac{1}{2v_{n0}} \frac{m_r}{m_\star}v_r^2,
\end{equation}
as shown in the corresponding cases [see Figs. \ref{fig:figure4}(c) and \ref{fig:figure4}(d)]. 
Note that $\Delta E_n$ can be up to about 0.24 MeV due to a larger $\Delta V_n$ (not shown) for the magnetized case with $B_0=50$ T, leading to further broadening of neutron spectrum due to $ \Delta E_n \propto  \Delta V_n ^2$.

In conclusion, driven by plasma instabilities, significant amplification of electromagnetic turbulence and the emergence of a non-Maxwellian ion distribution are observed in counter-streaming plasmas. Through theoretical analysis and numerical simulations, we have demonstrated that as the applied magnetic field perpendicular to the flows increases, both the growth rates and turbulence fields intensify. The ion distribution function exhibits a distinct super-Gaussian shape due to the competition between stochastic acceleration of turbulence and collisional thermalization. This ion kinetics is manifested in the neutron yields and spectra. Notably, the thermonuclear neutron yield is enhanced by more than 1 order of magnitude in the CD/CH case with $B_0 = 50 \ \mathrm{T}$, and the neutron spectra exhibit significant broadening in the magnetized CD/CD scenarios. Our results provide a good explanation for recent experimental observations in unmagnetized plasmas. We anticipate that the findings on magnetized plasmas will be validated through current astrophysical experiments utilizing available laser facilities. These insights offer profound implications for the amplification of turbulence fields and the generation of energetic particles in laboratory astrophysics \cite{BohdanPRL_TurbAmp,Bell_TrubAmp,PetersonPRL2021,Peterson_2022,Comisso_PRL2018,ZhdankinPRL2017,Petrosian_2012,ZhdankinPRL2019}.

J. Zhang, D. Wu, and D. W. Yuan initiated this work. P. Liu and D. Wu performed the simulations and data analysis. P. Liu and D. Wu drafted the manuscript. All the authors contributed to physical analysis.

We would like to thank the anonymous referees for their constructive comments and suggestions. This work was supported by the Strategic Priority Research Program of Chinese Academy of Sciences (Grant No. XDA250050500), the National Natural Science Foundation of China Grants (No. 12075204, No. 11875235, No. 61627901 and No. 11873061), the Shanghai Municipal Science and Technology Key Project (No. 22JC1401500), the Chinese Academy of Sciences Youth Interdiscipline Team (JCTD-2022-05), and the National Supercomputing Tianjin Center Fusion Support Program. D. Wu thanks the sponsorship from Yangyang Development Fund.


\bibliography{Reference}

\end{document}